\documentclass{aastex631}
\usepackage{soul}
\usepackage{CJK}

\shorttitle{Modulation of Antiprotons}

\shortauthors{Aslam et al.}
\graphicspath{{./}{figures/}}

\begin{document}
\begin{CJK*}{UTF8}{gbsn}

\title{Modulation of cosmic ray anti-protons in the heliosphere: simulations for a solar cycle}
\author[0000-0001-9521-3874]{O.P.M. Aslam}
\affiliation{Shandong Institute of Advanced Technology (SDIAT)\\
250100 Jinan, Shandong Province \\
People's Republic of China}

\author[0000-0003-0793-7333]{M.S. Potgieter}
\affiliation{Shandong Institute of Advanced Technology (SDIAT)\\
250100 Jinan, Shandong Province \\
People's Republic of China}
\affiliation{Institute for Experimental and Applied Physics (IEAP)\\
Christian-Albrechts-University in Kiel\\
24118 Kiel, Germany}

\author[0000-0002-4508-6042]{Xi Luo (罗熙)}
\affiliation{Shandong Institute of Advanced Technology (SDIAT)\\
250100 Jinan, Shandong Province \\
People's Republic of China}

\author[0000-0001-5844-3419]{M.D. Ngobeni}
\affiliation{Centre for Space Research, North-West University \\
2520 Potchefstroom, South Africa}
\begin{abstract}
The precision measurements of galactic cosmic ray protons from PAMELA and AMS are reproduced using a well-established 3D numerical model for the period July 2006 - November 2019. The resulting modulation parameters are applied to simulate the modulation for cosmic antiprotons over the same period, which includes times of minimum modulation before and after 2009, maximum modulation from 2012 to 2015 including the reversal of the Sun's magnetic field polarity, and the approach to new minimum modulation in 2020. Apart from their local interstellar spectra, the modulation of protons and antiprotons differ only in their charge-sign and consequent drift pattern. The lowest proton flux was in February-March 2014, but the lowest simulated antiproton flux is found to be in March-April 2015. These simulated fluxes are used to predict the proton to anti-proton ratios as a function of rigidity. The trends in these ratios contribute to clarify to a large extent the phenomenon of charge-sign dependence of heliospheric modulation during vastly different phases of the solar activity cycle. This is reiterated and emphasized by displaying so-called hysteresis loops. It is also illustrated how the values of the parallel and perpendicular mean free paths, as well as the drift scale, vary with rigidity over this extensive period. The drift scale is found to be at its lowest level during the polarity reversal period, while the lowest level of the mean free paths are found to be in March-April 2015.
\end{abstract}
\keywords{Particle astrophysics (96) --- Cosmic rays (329) 
--- Heliosphere (711) ---  Solar cycle (1487) --- Quiet Sun (1322)}
\section{Introduction} \label{sec:intro}
The level and extent of the occurrence of charge-sign dependent modulation in the heliosphere has always been a puzzling aspect because of the continuously varying nature and complexity of the Sun's activity cycles. This has been exasperated by the 
unavailability of continuous and long-term observations of galactic cosmic rays (GCRs) of opposite charge-signs with good precision and over an extended range of rigidity. This aspect, in particular, has changed markedly with the deployment of space borne experiments such as PAMELA, the Payload for Antimatter Matter Exploration and Light-nuclei Astrophysics \citep{2017NCimR..40..473P} in 2006 and AMS, the Alpha Magnetic Spectrometer \citep{2021PhR...894....1A} on the International Space Station in 2011. Before these missions, 
charge-sign dependent modulation was observed by the Ulysses mission \citep{1999GeoRL..26.2133H,2002JGRA..107.1274H,2003GeoRL..30.8032H}, which explored the heliosphere for the first time at high helio-latitudes, including two solar minima and a solar maximum period but separate measurements of electrons and positrons could not be made; see \url{https://sci.esa.int/web/ulysses} and the review on GCR observations by \cite{2006SSRv..127..117H, 2013SSRv..176..265H}. The two Voyager missions could also not distinguish between electrons and positrons but contributed significantly to our knowledge of the outer regions and boundary of the heliosphere, from the termination shock (TS) to the heliopause (HP), and when crossing the latter provided the first direct measurements of what can be considered the very local interstellar spectra (VLIS's) of various GCR species  \citep{2013Sci...341..150S,2019NatAs...3.1013S,2016ApJ...831...18C}, also see \url{https://voyager.jpl.nasa.gov/}. However, these reported VLIS's do not include those for positrons and antiprotons. For a recent review on the Voyager observations, and on other related GCR observations, see \citet{2022SSRv..218...42R}; for summaries of various aspects of charge-sign dependent observations by PAMELA, see e.g., \citet{2017ICRC...35.1091B,2017ApJ...834...89D,2017ICRC...35...12M}.

Even for the PAMELA and AMS missions reporting positron spectra and particularly anti-proton spectra over short time intervals and over a wide rigidity range remain an effort. Knowledge of the long-term modulation of these particular GCRs over solar cycles is therefore not at the same level than for protons and most other GCRs. However, numerical modulation models have become quite sophisticated and can easily be utilized to study the long-term modulation of these particles. Such a study for positrons was performed by \citet{2019ApJ...873...70A} who discussed the challenge of what to assume for the positron VLIS and what exactly to use as modulation parameters over the long-term. Following their approach, we simulated the proton spectra first and then study the modulation of anti-protons from one solar minimum to the next, including a period of the reversal of the magnetic field direction of the Sun. We find insightful results which are useful in making conclusions about the extent of charge-sign dependence over a complete solar cycle. For this we utilize the reported galactic proton measurements from PAMELA, combined with that from AMS, for the period from July 2006 to November 2019 \citep[][]{2013ApJ...765...91A,2018PhRvL.121e1101A, 2018ApJ...854L...2M}. 

In what follows, we describe the specific objectives, the utilized long-term observations, our modulation modelling approach and assumptions made, the model itself, the numerical solutions (results) for protons and anti-protons and the calculated corresponding flux ratios over the mentioned period. 
\section{Objectives and Observations} \label{sec:Observations}
Our first objective is to simulate and reproduce GCR proton spectra at the Earth. We utilize Carrington rotation averaged spectra for the period July 2006 - May 2011 \citep{2013ApJ...765...91A} and from January 2010 - February 2014 \citep{2018ApJ...854L...2M} over the kinetic energy range 80 MeV to 50 GeV measured by PAMELA. For a detailed description of the PAMELA detector and mission, see e.g., \cite{2017NCimR..40..473P}. It is noted that the highest proton spectra ever measured were reported by \citet{2013ApJ...765...91A} for the end of 2009. In this context, see also \citet{2014SoPh..289..391P,2021AdSpR..68.2953K}. 

We further utilize proton spectra from AMS from May 2011 - May 2017 (BRs 2426-2506) as published by \citet{2018PhRvL.121e1101A} and also for May 2011 - November 2019 (BRs 2426-2540) as published by \citet{2021PhRvL.127A1102A}. For 
the later period, up to November 2019, BR averages are calculated from the   reported daily resolution measurements over the rigidity range 1 - 100 GV.
For a detailed description about the AMS experiment, see e.g., \citet{2021PhR...894....1A}. Combining the AMS and PAMELA proton spectra over the period July 2006 - November 2019 provides us with more than 13 years of proton spectra as basis for our observational motivated and data driven simulations. 

Our second objective is to simulate the corresponding antiproton modulation for the period mentioned above. In order to accomplish this, all the modulation parameters found with the numerical model in reproducing the whole set of proton spectra are used unchanged to simulate the antiproton spectra for the exact same BRs. The only difference between the modulation of protons and antiprotons in the heliosphere is their drift patterns, oppositely directed during the same polarity phase, and, of course, their different VLIS's which are discussed in section \ref{sec:LIS}. 

A third objective is to use the simulated proton and anti-proton spectra to calculate their ratios at various rigidity values to establish how they change over time. Detailed time variations of the proton flux at 1.0 - 1.16 GV over the mentioned periods are explored. This provides insight of how particle drift and consequently charge-sign dependent modulation evolves over time with vastly different phases of the solar cycle.
\section{Solar activity proxies} \label{sec:intrinsic}
The continuously varying nature of the Sun's activity \citep{2013LRSP...10....1U,2015LRSP...12....4H} creates an 11-year time-dependence in the modulation of GCRs, and with the Sun's magnetic field changing its direction during each solar maximum, a 22-year cycle becomes evident in how GCRs are modulated. In this context, the Sun's magnetic 'polarity' is named positive (A $>$ 0) when the field lines are outward in its northern hemisphere and inward in the southern hemisphere. When the field lines change to be outward in the southern hemisphere and inward in the northern hemisphere, the 'polarity' is named negative (A $<$ 0). During e.g., the positive phase, positively charged GCRs are assumed to drift towards the Earth mainly, but not exclusively, through the polar regions of the heliosphere whereas negatively charged GCRs drift inwards mostly through the equatorial regions for the heliosphere. They then encounter the wavy heliospheric current sheet (HCS). See the reviews by e.g., \citet{2013SSRv..176..265H,2013SSRv..176..391K} and \citet{2014AdSpR..53.1415P,2017AdSpR..60..848P}.

During Solar Cycle 24, the change from $A<0$ to $A>0$ occurred from November 2012 - March 2014. This reversal is considered a relatively slow process and get completed in $\sim$16 months, see \citet{2015ApJ...798..114S}. For our modeling purposes we consider this reversal as a period with no well-defined polarity in neither the northern nor the southern hemisphere of the Sun. 

The time (solar) related activity is introduced to the model by using the tilt angle, $\alpha$, of the HCS along with the magnitude, $B$, of the heliospheric magnetic field (HMF) at the Earth as external parameters; both are considered good proxies for changing solar activity levels.
\begin{figure*}
\gridline{\fig{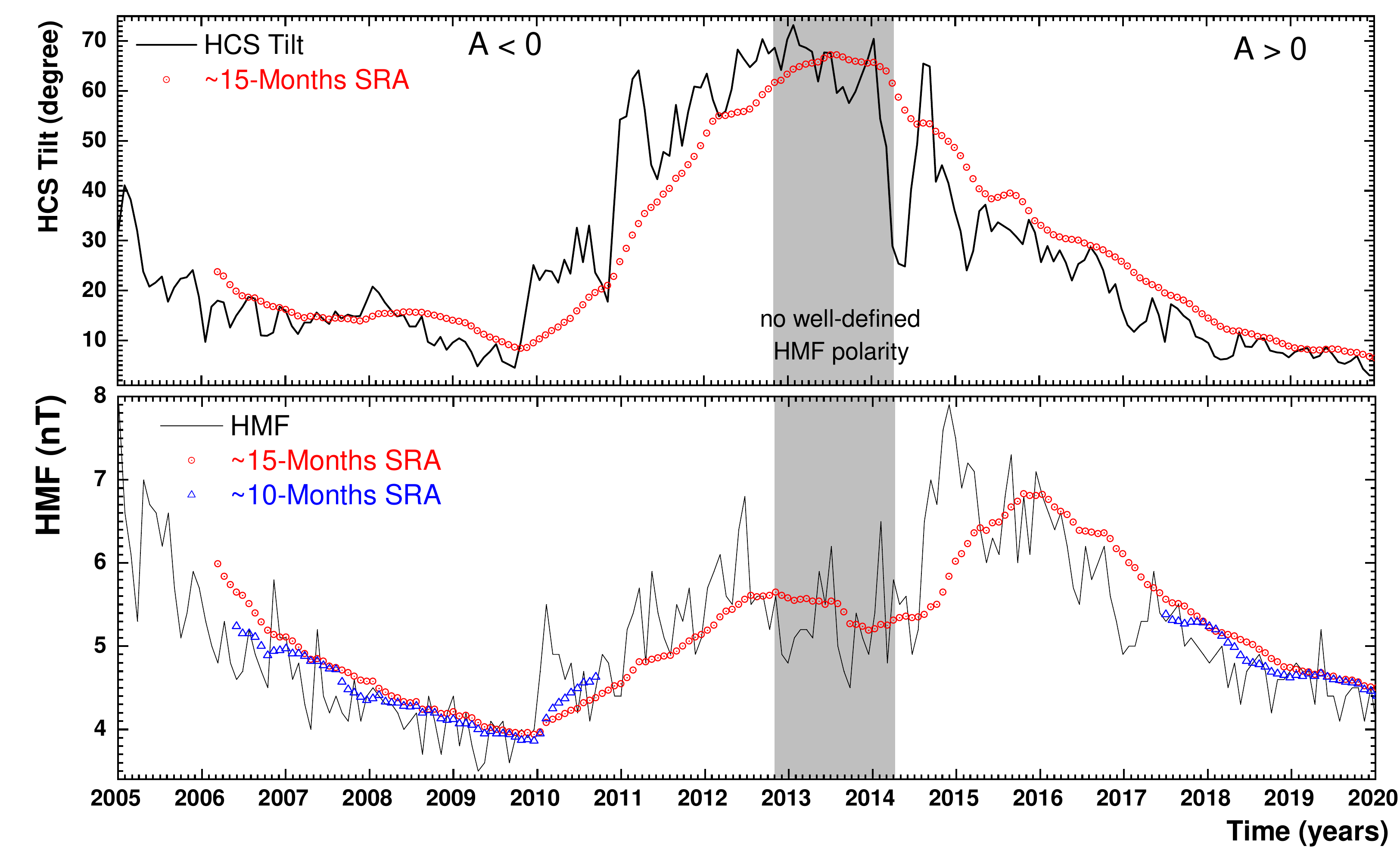}{0.9\textwidth}{}}
\caption{Lower panel: Magnitude of the HMF observed at the Earth averaged over Bartels Rotations (BR), adopted from \url{http://omniweb.gsfc.nasa.gov} for January, 2005 to December, 2019 (BRs 2340 to 2542) shown as solid line, together with $\sim$15-months (17-BRs) running averages, shown as red circles and $\sim$10-months (11-BRs) running averages, shown as blue circles.
Upper panel: Tilt angle ($\alpha$) of the Heliospheric Current sheet (HCS) as Carrington Rotation (CR) averages adopted from \url{http://wso.stanford.edu/} for the same period (CRs 2025 to 2226) shown as a solid line together with  $\sim$15-months (17 CRs) running averages shown as red circles.
Shaded parts indicate the period with no well-defined magnetic field polarity  (November 2012 - March 2014), when changing cycles from $A<0$ to $A>0$.
\label{fig1}}
\end{figure*}
In Figure \ref{fig1}, the variation of $\alpha$ is shown in the top panel from January 2005 - December 2019, averaged over Carrington Rotations from \url{http://wso.stanford.edu/}, along with calculated $\sim$15-months (17 rotations) running averages as implemented in the numerical model. In the lower panel, the observed $B$ at the Earth are averaged over Bartels Rotations (BRs), from \url{http://omniweb.gsfc.nasa.gov}, together with calculated $\sim$15 months (17 BRs) running averages and $\sim$10 months (11 BRs) running averages, as implemented in the numerical model. These averages reflect the assumed propagation times of these entities from the Earth to the HP.
For the HMF, the $\sim$10-month-averages are used for the solar minimum of July 2006 - October 2010 and again for August 2017 - November 2019, when the averaged $\alpha$ is below 20$^{\circ}$, while the $\sim$15-month-averages are used for other periods. Selecting these running averages to be used as solar activity proxies in our modeling approach is explained and discussed by \citet{2015ApJ...810..141P, 2019ApJ...873...70A, 2021ApJ...909..215A}. 

Another solar activity related feature changing with time in the model is the heliolatitude dependence of the solar wind speed on a global scale. In accordance with Ulysses observations, it is assumed that it changes from 430 km s$^{-1}$ in the equatorial plane to 750 km s$^{-1}$ in the polar regions during solar minimum activity conditions, see \citet{2002GeoRL..29.1290M,2006SSRv..127..117H}.
As the Solar Cycle progresses towards maximum, this latitudinal dependence gradually decreases in the polar regions to become 450 km s$^{-1}$ when $\alpha$ is 65$^{\circ}$ or higher.

The width of the inner heliosheath affects the modulated GCR intensity at the Earth \citep{2016SoPh..291.2181V}. The TS moves inward and outward in response to changes in the dynamic pressure of the solar wind, and that changes the width of the inner heliosheath, as reported by \citet{2011ApJ...734L..21R} based on the Voyager 2 measurements. This dynamic nature of the TS is incorporated in the model, in addition to changing $\alpha$ and the HMF magnitude; for similar approaches, see \citet{2004JGRA..109.1103L,2005AdSpR..35.2084L,2014SoPh..289.2207M}. In the model, the position of the TS is changed gradually from 88 au (July 2006) to 80 au (December 2009), gradually increasing back to 88 au by April - May 2014, and then gradually decreasing to 78 au by mid-2019; see Figure 3 by \citet{2016SoPh..291.2181V}. The spherically shaped HP is assumed to be at a radial distance of 122 au from the Sun.

\section{Modeling Approach} \label{sec:modeling}

Once inside the heliosphere GCRs are subjected to an outward convection by the solar wind, adiabatic deceleration due to the expanding solar wind, diffusion along and across the turbulent HMF, and gradient, curvature and HCS drifts in the global HMF. These processes are responsible for altering the differential intensities of GCRs as a function of energy (rigidity), position in the heliosphere and time (solar activity). We simulate the transport of GCRs within the heliosphere by including all of these modulation processes into a full three-dimensional (3D) numerical model to compute the differential intensity of GCRs at different radial distances from the Earth up to the HP and spectra at the Earth over the rigidity range 10 MV - 70 GV. The VLIS's are specified as an initial condition (unmodulated spectra) at the HP. The predicted small modulation beyond the HP, as simulated by e.g., \citet{2011ApJ...735..128S,2015ApJ...808...82L}, is neglected.
\subsection{Modulation model} \label{sec:model}
The model is based on the numerical solution of Parker's transport equation \citep{1965P&SS...13....9P} which follows the basic transport equation (TPE) for charged particle motion in large and small scale fluctuating magnetic fields, and averaged over the pitch and phase angles of propagation particles, with the assumption that GCRs are approximately isotropic. For the heliosphere this TPE is described as:  
\begin{equation}
\frac{\partial f}{\partial t} = - \vec{V}_{sw} \cdot \nabla \it{f} - \langle \vec {v}_{D} \rangle \cdot \nabla \it{f} + \nabla \cdot (\bf{K}_{s} \cdot \nabla \it{f}) + \frac {1}{3} (\nabla \cdot \vec{V}_{sw}) \frac {\partial f} {\partial \ln p} 
\label{Eq1}
\end{equation}
where $f (\vec {r}, p, t)$ is the omnidirectional GCR distribution function, $\it{p}$ is momentum, $\it{t}$ is time, and $\vec {r}$ is the vector position in 3D. If $\it{P}$ is particle rigidity, the differential intensity is given by $\it{P}^{2}f$ ($\it{p}^{2}f$). The first two terms shown on the right-hand side of this equation represent outward convection caused by the expanding solar wind with velocity ($\vec {V}_{sw}$) and the averaged particle drift velocity $\langle \vec {v}_{D} \rangle$ (pitch angle averaged guiding center drift velocity) i.e. particle drift caused by the HMF's gradients, curvatures and the wavy HCS. 
The third term represents spatial diffusion described by the symmetry diffusion tensor $\bf{K}_{s}$. When the TPE is written in heliocentric spherical coordinate system, the three coordinates $\it{r}$, $\theta$, and $\phi$ represent radial distance, polar and azimuth angles, respectively. According to this system, the Earth is located at 1 au in the equatorial plane with $\theta = 90^{\circ}$. 

In order to solve the TPE, six elements in this tensor related to diffusion are to be specified: $K_{rr}$, $K_{r\phi}$, $K_{\theta \theta}$, $K_{\phi r}$, and $K_{\phi \phi}$. 
These five diffusion coefficients in terms of diffusion parallel and perpendicular to the HMF direction are as follows: 
\begin{equation}        
K_{rr} = K_{\parallel} {\cos}^{2} \psi + K_{\perp r} {\sin}^{2} \psi. 
\label{Eq2}
\end{equation}
\begin{equation} 
K_{\theta \theta} = K_{\perp \theta}.
\label{Eq3}
\end{equation}
\begin{equation} 
K_{\phi \phi} = K_{\perp r} {\cos}^{2} \psi + K_{\parallel}{\sin}^{2} \psi. 
\label{Eq4}
\end{equation}
\begin{equation} 
K_{\phi r} = (K_{\perp r} - K_{\parallel}) {\cos} \psi {\sin} \psi = K_{r \phi}. 
\label{Eq5}
\end{equation}
This means, for example, that the effective radial diffusion coefficient $K_{rr}$ is a combination of the parallel diffusion coefficient ($K_{\parallel}$) and the radial perpendicular diffusion coefficient ($K_{\perp r}$), with $\psi$ the spiral angle of the background HMF; $K_{\theta \theta}$ = $K_{\perp \theta}$ is the effective perpendicular diffusion coefficient in the polar direction; $K_{\phi \phi}$ describes the effective diffusion in the azimuth direction and $K_{\phi r}$ diffusion in the $\phi r$-plane.

The fourth term in the right hand side of Equation \ref{Eq1} represents adiabatic energy changes which depends on the sign of the divergence of $\vec {V}_{sw}$. If $\nabla \cdot \vec{V}_{sw}>0 $, adiabatic energy loss occurs as is the case in most of the heliosphere, except perhaps inside the heliosheath where we assume that $\nabla \cdot \vec{V}_{sw} = 0$; see also \citet{2006ApJ...640.1119L}. Adiabatic energy changes becomes dominant at lower rigidity in the case of modulated spectra for protons, antiprotons and GCR isotopes; see illustrations of this process by \citet{1982Ap&SS..84..519M,2011JGRA..11612105S}.

The drift velocity can be written as follows:
\begin{equation}
\langle \vec {v}_{D} \rangle = \nabla \times K_{D} \frac{\vec B}{B},
\label{Eq6}
\end{equation}
where $K_{D}$ is the generalized drift coefficient, and $\vec{B}$ is the HMF vector with magnitude $\it{B}$. 
When Equation \ref{Eq1} is rewritten not showing the drift velocity explicitly, then a generalized propagation tensor replaces the diffusion tensor and four elements must then be added to describe gradient and curvature drift: $K_{r\theta}$, $K_{\theta r}$, $K_{\theta \phi}$, and $K_{\phi \theta}$. The three components of the drift velocity (in the radial, polar and azimuth directions) in terms of these elements of the propagation tensor are then given by:
\begin{equation}
   \langle \vec {v}_{D} \rangle_{r} = -\frac{A}{r \sin\theta} \frac{\partial}{\partial \theta} (\sin\theta K_{\theta r}), \\ 
   \langle \vec {v}_{D} \rangle_{\theta} = -\frac{A}{r} \left[\frac{1}{\sin\theta} \frac{\partial}{\partial \phi} (K_{\phi\theta}) + \frac{\partial}{\partial r} (r K_{r \theta}) \right], \\
   \langle \vec {v}_{D} \rangle_{\phi} = -\frac{A}{r} \frac{\partial}{\partial \theta} (K_{\theta \phi}),
\label{Eq7}
\end{equation}
where $A$ is a constant that denotes the HMF polarity. In essence, it determines the drift directions of charged particles:
\begin{equation}
    A = +1 (qA > 0); A = -1 (qA < 0),
\label{Eq8}
\end{equation}
with $q$ the particle's charge-sign.

Evidently, the geometry and magnitude of the HMF are important for the modulation process, particularly in the case of particle drift. We start with a straight-forward HMF as described by \citet{1958ApJ...128..664P}: 
\begin{equation}
\vec {B} = B_{0}A\Bigg[\frac{r_{0}}{r}\Bigg]^{2}(\hat {e}_{r}-\tan\psi \hat{e}_{\phi}),
\label{Eq9}
\end{equation}     
where $\hat {e}_{r}$ and $\hat {e}_{\phi}$ are unit vectors in the radial and azimuth directions;
$B_{0}$ is the magnitude of the HMF at the position $r_0$, in our approach assumed to be at the Earth, and
\begin{equation}
  {\tan}  \psi = \Omega \frac {(r - r_{\odot})} {V_{sw}} \sin\theta, 
  \label{Eq10}
\end{equation}     
where $r_{\odot}$ is the solar radius (0.005 au) and $\Omega$ is the average angular rotation speed of the Sun (2.66 $\times$ 10$^{-6}$ rad s$^{-1}$). The magnitude of this Parkerian HMF is 
\begin{equation}
B = B_{0}\Bigg[\frac{r_{0}}{r}\Bigg]^{2}\sqrt{1 + (\tan \psi)^{2}.}
\label{Eq11}
\end{equation} 
However, this HMF is modified as proposed by 
\citet{1991ApJ...370..435S} based on their insight that the magnetic field spirals are relatively more tightly wound than that predicted by the original Parker theory. They argued that the differential rotation of the Sun would cause small azimuthal magnetic field components to develop and that would lead to larger spiral angles at larger radial distances. They proposed a modification so that the expression for the HMF spiral angle becomes:
\begin{equation}
\tan \psi = \frac {\Omega (r - r_{b}) \sin\theta} {V_{sw} (r, \theta)} - \frac {r V_{sw} (r_{b}, \theta)} {r_{b} V_{sw} (r, \theta)} \Bigg( \frac {B_{T}(r_{b})}{B_{R}(r_{b})}\Bigg)
\label{Eq12}
\end{equation}
where $B_{T}(r_{b})/ B_{R}(r_{b})$ is the ratio of the azimuthal to the radial magnetic field components at a position $r_{b}$ near the solar surface. 
It is argued that the ratio $B_{T}(r_{b})/ B_{R}(r_{b})$ $\approx$ -0.02 at a position $r_{b}$ = 20$r_{\odot}$. With $r_{\odot}$ = 0.005 au as the solar radius, the value $r_{b}$ = 20$r_{\odot}$ and the ratio $B_{T} (r_{b})/B_{R} (r_{b})$ = -0.02 are constants that determine the HMF modification. This modification keeps the basic Parkerian geometry (and $\nabla \cdot \vec{B}=0$) but modifies its magnitude progressively toward the poles of the heliosphere, which essentially give less effective particle drift in the polar regions of the heliosphere. The motivation for this modification from a modulation modelling point of view is discussed by \citet{2013LRSP...10....3P,2015ApJ...810..141P,2017A&A...601A..23P}. An elaborate discussion and illustration of its relevance and effects on GCR modulation is given by \citet{2015Ap&SS.360...56R}.

The functional forms of the three diffusion coefficients ($K_{\parallel}$,$K_{\perp r}$,$K_{\theta \theta}$) and the drift coefficient ($K_{D}$) as used in this model are given and discussed in section \ref{sec:Drift}. 
\begin{figure}[ht!]
\gridline{\fig{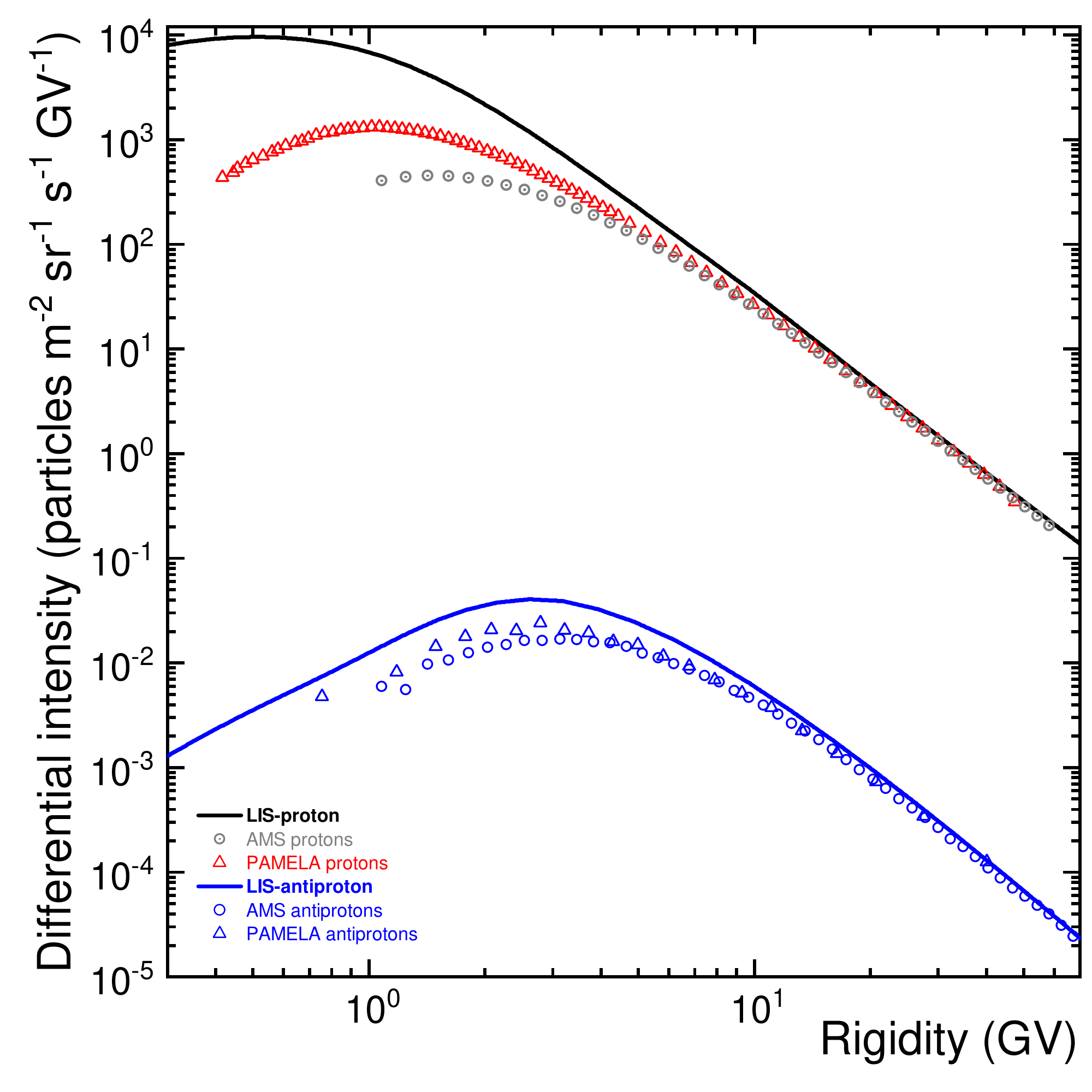}{0.8\textwidth}{}}
\caption{Very local interstellar spectra (VLIS's) at the HP (122 au) as a function of rigidity: for GCR protons as the black solid line from \citet{2019ApJ...878...59B} and for anti-protons as the blue solid line from \citet{2021Potgieter}. Observations at the Earth are shown as red triangles indicating PAMELA protons, averaged for July 2006 - January 2010 \citep{2013ApJ...765...91A}, with blue triangles indicating PAMELA antiproton averages for the same period \citep{2013JETPL..96..621A}; blue circles are AMS antiproton averages for May 2011 - May 2015 \citep{2016PhRvL.117i1103A} and the gray circles are AMS protons for the same period \citep{2018PhRvL.121e1101A}.} 
\label{fig2}
\end{figure}
\citet{2014SoPh..289..391P,2015ApJ...810..141P} described the global features of the  model used here in detail e.g. the expression for the full TPE in spherical coordinates; for the expression of the global solar wind structure, see their equations 12 and 5, respectively. \citet{2016AdSpR..57.1965R} give the expression for the wavy HCS as used here. Details are also discussed by \citet{2015ApJ...810..141P,2017A&A...601A..23P} and the same or a similar model was applied by  \citet{2019ApJ...873...70A,2021ApJ...909..215A,2019ApJ...871..253C,2020Ap&SS..365..182N,2022AdSpR..69.2330N}. 
Comprehensive reviews of this modeling approach, and underlying theory for the global modulation of GCRs, are given by \citet{2014AdSpR..53.1415P,2017AdSpR..60..848P} and 
\citet{2021Potgieter}. 
As done for previous applications, we continue to assume a steady-state so that the model is not well suitable to study GCR transient events as explained by \citet{2021ApJ...909..215A}; the applied Alternating Direct Implicit (ADI) numerical method is not stable when a fifth numerical dimension is incorporated into the numerical scheme (that is, three spatial, one for energy dependence, and one for time dependence). If all three spatial dimensions are used, the time dependence must be sacrificed. The only alternative for 3D with a full time-dependence is to use numerical models based on Stochastic Differential Equations (SDEs) which is a totally different approach to modulation studies, see e.g. \citet{2012Koppetal,2015Dunzlaffetal,2019ApJ...878....6L,2020ApJ...899...90L,2021ApJS..257...48S,2021ApJS..256...18S}, and references therein. Many 2D time-dependent modeling results, based on an ADI numerical scheme, have been published over the years e.g., \citet{1992ApJ...397..686L,1995ApJ...442..847L,2003AnGeo..21.1359F,2004JGRA..109.1103L,2005AdSpR..35.2084L,2012AdSpR..49.1660N,2014SoPh..289.2207M}.
\subsection{Very Local Interstellar Spectra} \label{sec:LIS}
In order to reproduce observed proton spectra, as vindication of the model, and to simulate subsequently antiproton spectra, we have to specify for each type (specie) of GCR particles an input spectrum as an initial condition in the numerical model; it is done in the form of a VLIS at the HP.
\citet{2019ApJ...878...59B,2021PAN....84.1121B} illustrate the approach of how to construct self-consistent VLIS's for GCRs by utilizing available comprehensive galactic propagation models (GALPROP) along with a 3D modulation model and with relevant validating observations. At high enough rigidity where modulation may be considered negligible (\cite{2014AdSpR..53.1015S}), observations from AMS and PAMELA are used, whereas and at low rigidity for protons, observations from Voyager 1 and Voyager 2 beyond the HP are used \citep{2013Sci...341..150S,2019NatAs...3.1013S,2016ApJ...831...18C}.
See \citet{1998ApJ...509..212S} for information about GALPROP and a description of the numerical computation of the propagation of primary and secondary nucleons, and \citet{2002ApJ...565..280M,2005AdSpR..35..156M} especially for numerical computation of antiproton propagation. Studies determining VLIS's for modulation modeling purposes, have been reported by e.g., \citet{2015ApJ...815..119V,2016Ap&SS.361...48B,2016ApJ...829....8C,2019AdSpR..64.2459B, 2020ApJ...889..167B}.

The VLIS for galactic protons obtained in this way, is depicted in Figure \ref{fig2} in units of "particles m$^{-2}$ sr$^{-1}$ s$^{-1}$ GV$^{-1}$" as a function of rigidity over the range 0.30 - 70.0 GV. (The Voyager data are not repeated here in these units; see Figure 6 by \citealt{2021Potgieter}). The VLIS for antiprotons computed with the GALPROP code and then adjusted as described by \citet{2021PAN....84.1121B} and \citet{2021Potgieter}, is shown in comparison.  Observational spectra for protons and antiprotons at the Earth are shown for illustrative purposes and described in the figure's caption. 

Apart from showing the VLIS's applied in the modelling, this figure also illustrates the total modulation obtained with the model between the HP (122 au) and Earth (1 au) and how vastly different that is for antiprotons compared to protons. The peak intensity in the proton VLIS occurs at a much lower rigidity than for antiprotons, subsequently also for the modulated spectra and it shows that the additional modulation that protons experienced is markedly more than for antiprotons based on the illustrative averaged intensity in 2006 - 2009 to that in 2011 - 2015. These differences are explored and reported in what follows. 
\section{Results} \label{sec:results}
\begin{figure*}
\gridline{\fig{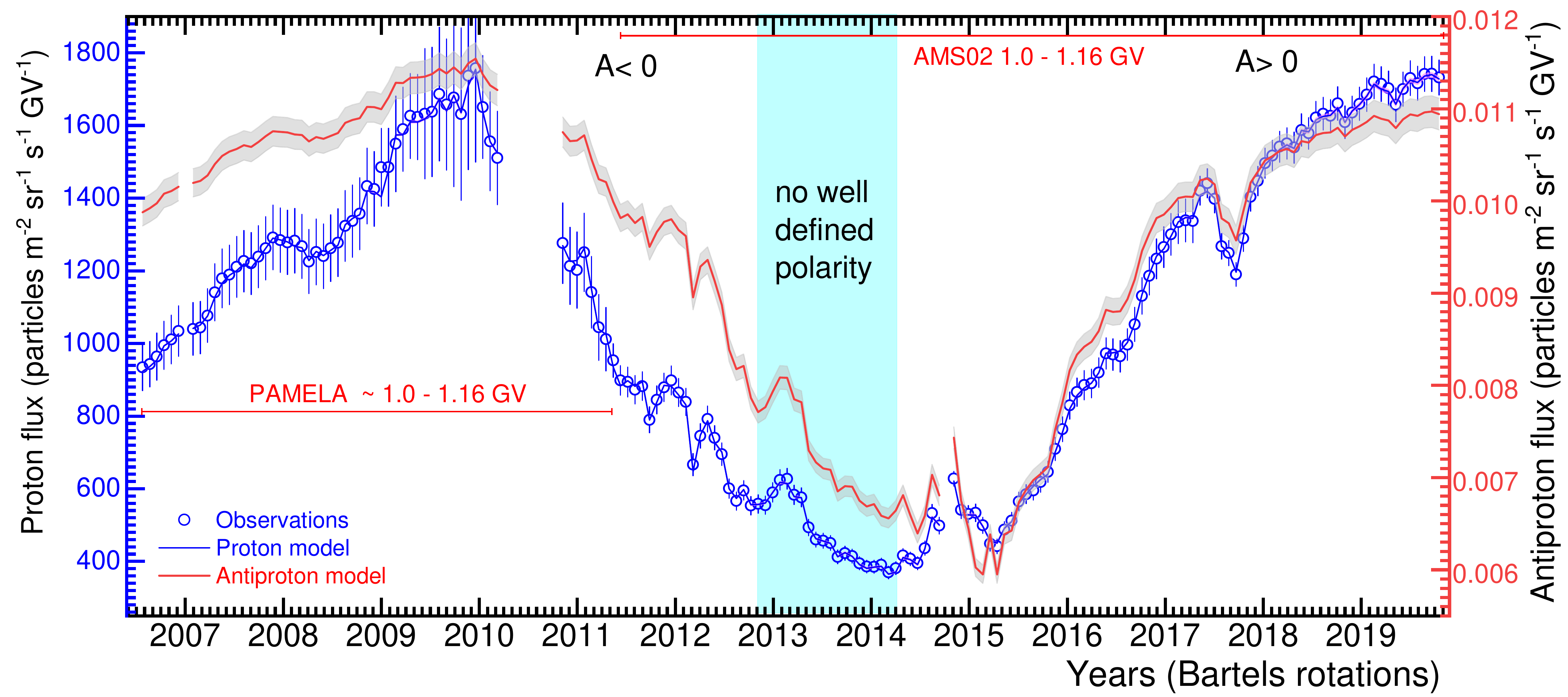}{0.95\textwidth}{}}
\caption{Simulated proton flux (blue solid line) from July 2006 - November 2019 (BRs 2361-2540) along with a standard error of mean (SEM; standard deviation /$\sqrt{n}$ ) for 1.0 - 1.16 GV in comparison with the observed proton flux (blue circles) of the same rigidity. For July 200 - May 2011 observations are from PAMELA \citep{2013ApJ...765...91A, 2018ApJ...854L...2M} and for May 2011 - November 2019  from AMS \citep{2018PhRvL.121e1101A, 2021PhRvL.127A1102A}. 
Corresponding simulated antiproton flux is shown (red solid line) along with its SEM (gray shading). No published BRs averaged antiproton observations are available for comparison. Shaded range indicates the period without a well-defined HMF polarity (November 2012 - March 2014) as it changes from $A<0$ to $A>0$.
\label{fig3}}
\end{figure*}

We first simulate and reproduce proton spectra for each Carrington Rotation (from PAMELA) or Bartels Rotation (from AMS) from July 2006 - November 2019 over a wide range of rigidity. Based on these spectra, the differential intensity for the rigidity bin 1.0 GV - 1.16 GV is shown over time in Figure \ref{fig3} and compared with observations for the same rigidity bin. Once the observed proton spectra are reproduced satisfyingly, the same set of modulation parameters is used to simulate the antiproton spectra for the same period of time and rigidity as shown in Figure \ref{fig3}. Corresponding published antiproton measurements are not available for comparison.

Inspecting the time trends in Figure \ref{fig3}, it is noticed how both the proton and antiproton fluxes show a gradual increase from July 2006 onward to reach the highest level by the end of 2009. The proton flux shows $\sim$75\% increase over this period (based on the averaged value over the whole 2006 - 2019 period), but the antiproton flux shows an increase of only $\sim$17\%. This different trend in the intensity-time profiles is considered to be indicative of the effect of different drift patterns of these oppositely charged particle.
After December 2009, both fluxes progressively decrease as the solar cycle progresses towards maximum, with the proton flux reaching its minimum level by the end of the polarity reversal phase (March - April 2014), but the antiproton flux reaches a first minimum somewhat just after the reversal ended  and then another minimum in February - March 2015. This is related to the peculiar behaviour of both the tilt angle and HMF in late 2014 as shown in Figure \ref{fig1}. The rate of decrease of the antiproton flux from December 2009 to March 2015 is different compared to protons, again considered to be indicative of the different drift patterns and consistent to drift model predictions that the negatively charged GCRs should have a flatter intensity-time profile during $A<0$ cycles. Before the recovery starts in March - April 2014, the proton flux then is only $\sim$20\% of the flux in December 2009.
The antiproton flux is reduced up to only $\sim$50\% of the December 2009 level before the recovery start.
After the HMF reversal, as the solar cycle progresses toward minimum, both the fluxes increase, with the proton flux reaching almost the same level in September-October 2019 as in December 2009. Over the same period the antiproton flux at this rigidity reaches a level of $\sim$95\% of the level of December 2009.

\subsection{Diffusion and particle drift parameters over a solar cycle
\label{sec:Drift}}
\begin{figure*}
\gridline{\fig{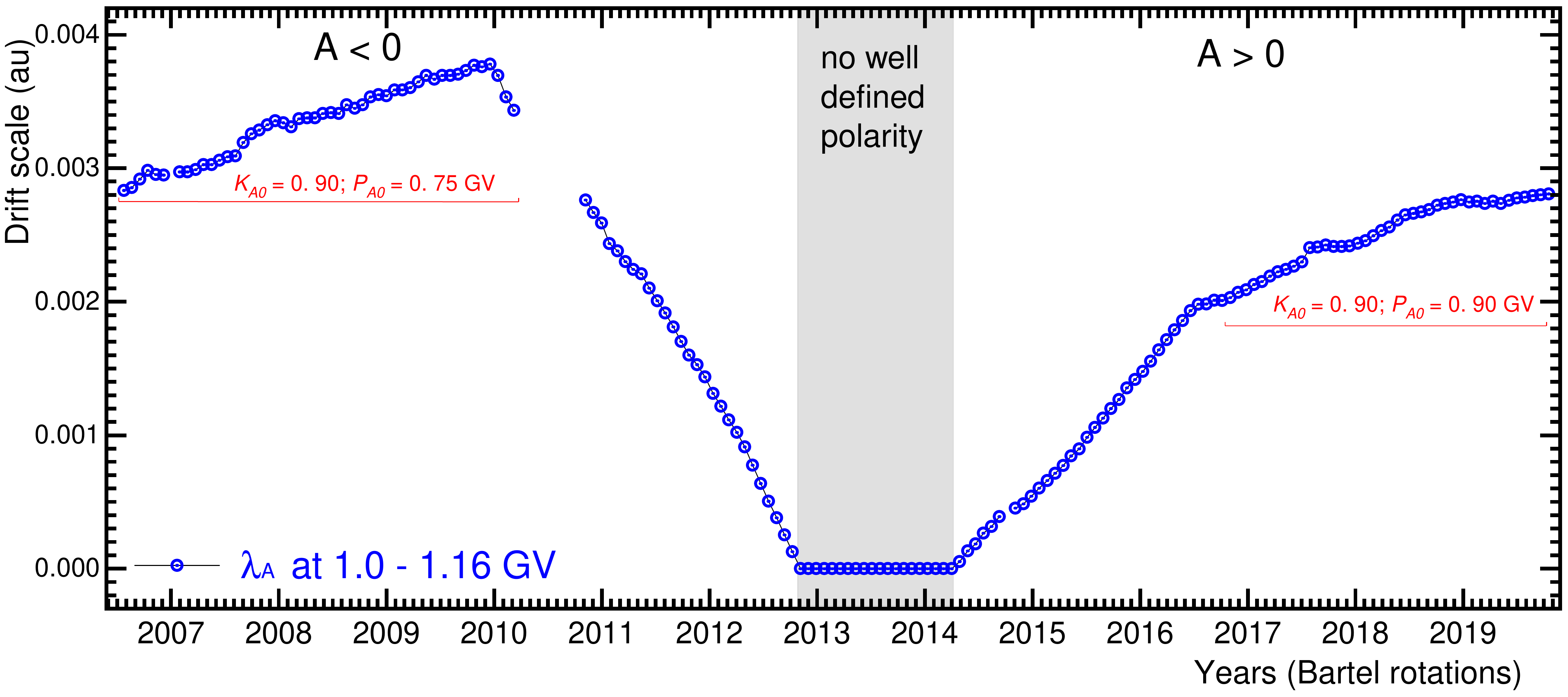}{0.9\textwidth}{(a)}
}
\gridline{\fig{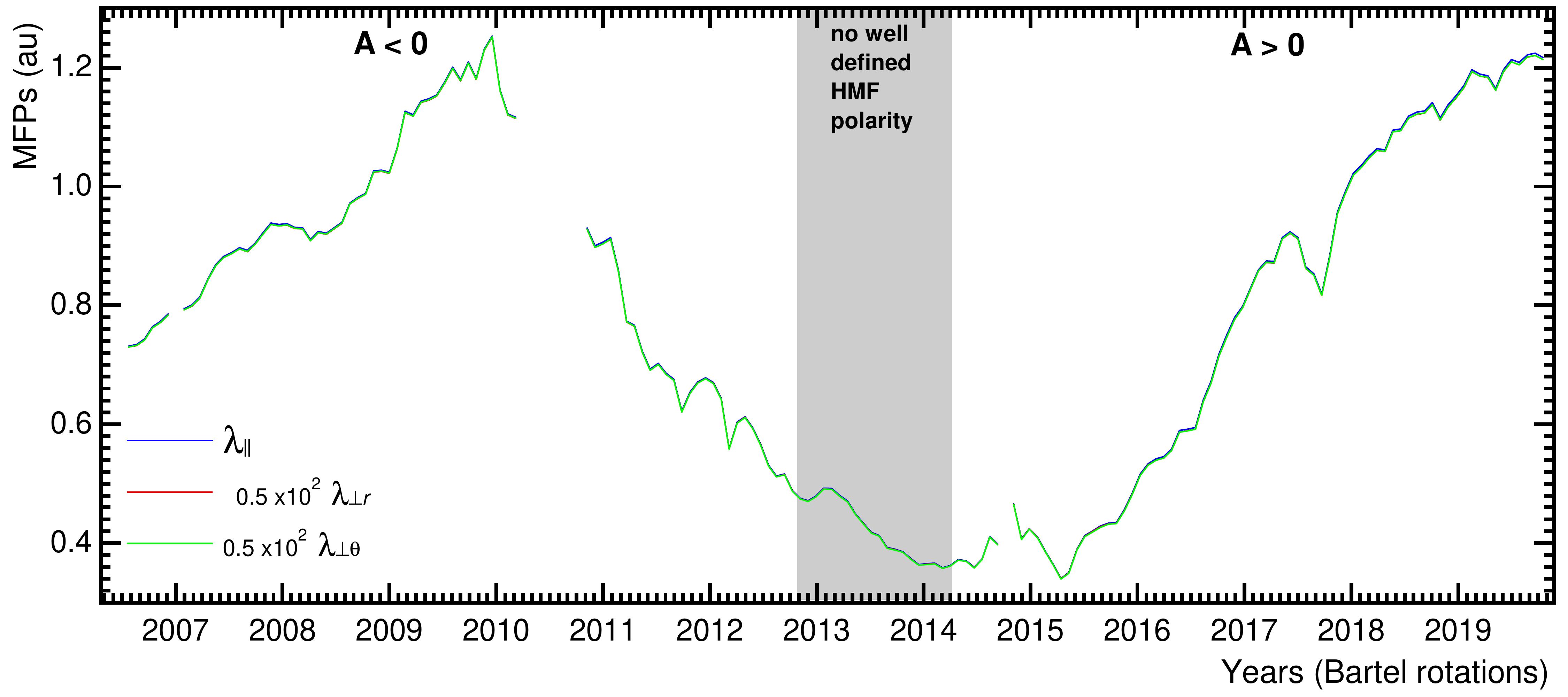}{0.9\textwidth}{(b)}
}
\caption{Panel (a): Variation of the drift scale over time at the Earth for 1.0-1.16 GV from July 2006 to October 2019.
Panel (b): Variation of the three MFPs $\lambda_{\parallel}$, $\lambda_{\perp r}$ and $\lambda_{\perp \theta}$ 
over the same period and rigidity. Both $\lambda_{\perp r}$ and $\lambda_{\perp \theta}$ are multiplied by $0.5\times10^{2}$ for a direct comparison with $\lambda_{\parallel}$ indicating that they have the exact same time dependence. In these panels the emphasis is on the trends over time from before the 2009 solar minimum to the next.
\label{fig4}}
\end{figure*}

\begin{figure}
\gridline{\fig{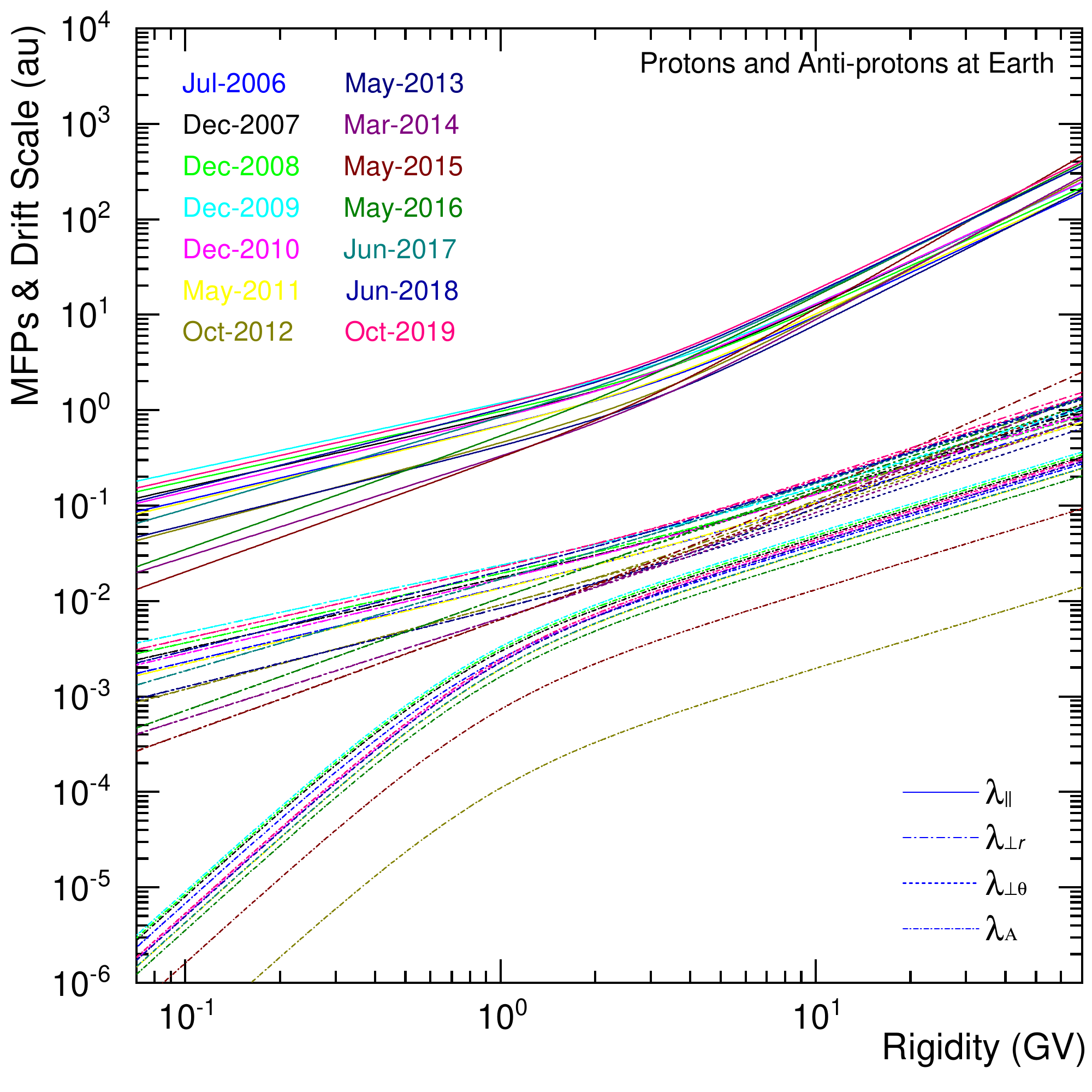}{0.60\textwidth}{}}
\caption{Rigidity dependence of the three MFPs ($\lambda_{\parallel}$, $\lambda_{\perp r}$, \& $\lambda_{\perp \theta}$) and drift scale ($\lambda_{A}$) at Earth for both protons and antiprotons (see Equations 15 to 19). It also shows how they vary from July 2006 - November 2019 with values for selected BRs indicated with different colors, e.g., July 2006 is when PAMELA observations began; December 2009 is when the highest proton intensity was observed; May 2011 is when the AMS observations began; October 2012 is end of the $A<0$ polarity cycle; March 2014 is when the lowest proton intensity was observed; May 2016 was when the drift scale recovery is completed.
\label{fig5}}
\end{figure}

In this section, we elaborate on the diffusion and drift coefficients used in the model to reproduce the proton spectra and to simulate the antiproton flux. These coefficients contain the essence of what determines the level of solar modulation over a solar cycle.
We emphasize first the variation of the drift scale, with results shown to vary over time in Figure \ref{fig4}(a) at a chosen rigidity, and then as a function of rigidity in Figure \ref{fig5}. The parallel and perpendicular mean free paths are shown in Figure \ref{fig4}(b) at a chosen rigidity over time, and also as a function of rigidity in Figure \ref{fig5}.     

The expression for a generalized drift coefficient, can be written as:
\begin{equation} 
K_{D} = \frac {\beta P} {3B_{m}} f_{D} = \frac {\beta P} {3B_{m}} \Bigg[ \frac {(\omega \tau)^{2}}{1+(\omega \tau)^{2}},\Bigg]
\label{Eq13}
\end{equation}
with $B_{m}$ the magnitude of the modified HMF (see section \ref{sec:model}) and 
where $\beta$ = $v/ c$ is the ratio of particle speed to the speed of light, and $\omega$ is the particle gyro-frequency with $\tau$ the average time between the scattering of GCR particles in the HMF. In most of the numerical modeling studies, it is assumed that $\omega \tau \gg 1$ in the heliosphere so that the drift coefficient takes its simplest form:
\begin{equation} 
K_{D} = \frac {\beta P}{3B_{m}},
\label{Eq14}
\end{equation}
known as weak scattering drift. Establishing $\tau$ is complicated, even controversial; an elaborate turbulence theory is required to understand how $\omega\tau$ could change as a function of rigidity and space throughout the entire heliosphere, and especially over the solar cycle; see e.g., \citet{2018ApJ...859..107M,2018ApJ...856...94Z,2021ApJ...908..167E} and the review by \citet{2022SSRv..218...33E}.
The term $f_{D}$ 
is called the drift reduction factor and is determined by how diffusive scattering is described; if $f_{D}$ = 0, then $K_{D}$ and therefore the drift velocity $\langle v_{D} \rangle$ becomes zero so that drift effects vanish from the modulation model to produce non-drift solutions; if $f_{D}$ approaches 1, drift is at a maximum, and $K_{D}$ then has the weak scattering value. Theoretically, the largest possible value for $K_{D}$ is obtained  under such conditions and when the Parker HMF is kept unmodified; for detailed discussions, see e.g., \citet{2015AdSpR..56.1525N}, and references there-in.
If this is the case, drift effects in GCR modulation are predicted to be very large and dominant as found in original numerical models by \citet[][where the $A>0$ and $A<0$ computed spectra were the opposite of what was observed]{1979ApJ...234..384J}, \citet{1981ApJ...243.1115J}, \citet{1983ApJ...265..573K} and \citet{1985ApJ...294..425P}, to mention only a few. 
The function $f_{D}$ is also used to adjust the rigidity dependence of $K_{D}$, which is the most effective direct way of suppressing drift effects at low rigidity as required by $\it{Ulysses}$ observations of latitudinal gradients; see reviews by \citet{2001SSRv...97..309H}; \citet{2006SSRv..127..117H}; \citet{2013SSRv..176..265H}. As such, Equation (\ref {Eq13}) gets a more practical form \citep[e.g.][]{2015AdSpR..56.1525N,2018ApJ...859..107M} and has been widely used in numerical drift models:
\begin{equation}
K_{D} = \frac {\beta P} {3B_{m}} f_{D} = K_{A0}  \frac {\beta P} {3B_{m}} \frac {(P/P_{A0})^{2}}{1+(P/P_{A0})^{2}}.
\label{Eq15}
\end{equation}           
Here, $K_{A0}$ is dimensionless and used to adjust $K_{D}$ in a pragmatic way; if $K_{A0}$ = 1.0, it is called 100\% drift (full weak scattering). In this study, we keep $K_{A0}$ = 0.90 for the July 2006 - June 2010 period, then gradually decreases it to a minimum level for the polarity reversal phase from November 2012 - March 2014. After the completed reversal, it is increased gradually back and by mid-2016 it is again 0.90, and kept this way until the end of the simulation period. It is emphasized that this type of variation is needed to reproduce the proton flux shown in Figure \ref{fig3}. Furthermore, the value of $P_{A0}$ = 0.75 GV from July 2006 - April 2010, but after November 2010 it is 0.90 GV. This parameter determines the rigidity value below which $K_{D}$ is reduced in terms of rigidity.
Adjusting $K_{A0}$ and $P_{A0}$ along with ${B_m}$ over time is the explicit way of drift adjustment; see also \citet{2013LRSP...10....3P,2014AdSpR..53.1415P}, and \citet{2016AdSpR..58..453N,2018Ap&SS.363..156N} who also discussed the implicit way.
The global drift scale, $\lambda_{A}$ = (${3}/{v}) K_{D}$, in units of astronomical units (au), is found to vary from July 2006 - November 2019 at 1.0 - 1.16 GV, which is selected as the lowest rigidity observed by AMS. This is plotted in Figure \ref{fig4}(a). The times when $K_{A0}$ and $P_{A0}$ are kept unchanged are shown so that the time trend in $\lambda_{A}$ is caused by the changes in HMF strength as shown in Figure \ref{fig1}.     
The drift scale as a function of rigidity at the Earth is shown in Figure \ref{fig5} as the lower set of graphs, also indicating how it changes with time (solar activity). The rapid decrease in $\lambda_{A}$ below rigidity $\sim$1 GV is the direct result of the scaling of $f_{D}$, essentially determined by $P_{A0}$ in Equation \ref{Eq15}; evidently, its value may also change with solar activity. 

Next, we focus on the expressions for the three diffusion coefficients, first, $K_{\parallel}$ is assumed to be:
\begin{equation}
K_{\parallel} = (K_{\parallel})_{0} \beta \Bigg(\frac {B_{u}}{B_m}\Bigg) \Bigg(\frac {P}{P_{u}}\Bigg)^{c_{1}} \left[ \frac {\Bigg(\frac {P}{P_{u}}\Bigg)^{c_{3}} + \Bigg(\frac {P_{k}}{P_{u}}\Bigg)^{c_{3}}}{ 1+ \Bigg(\frac {P_{k}}{P_{u}}\Bigg)^{c_{3}}} \right]^{\frac {c_{2 \parallel} - c_{1}}{c_{3}}}
\label{Eq16}
\end{equation}
where $(K_{\parallel})_{0}$ is a scaling constant in units of $10^{22}$ cm$^{2}$s$^{-1}$, with the rest of the equation written to be dimensionless with $P_{u} = 1.0$ GV, and $B_{u}$ = 1.0 nT (in order to preserve the units in cm$^{2}$ s$^{-1}$). Here $c_{1}$ is a power index that may change with time; $c_{2 \parallel}$ and $c_{2 \perp}$ together with $c_{1}$ determine the slope of the rigidity dependence, respectively, above and below a rigidity with the value $P_{k}$ which may also change with time; $c_{3}$ determines the smoothness of the transition. The rigidity dependence of $K_{\parallel}$ is thus a combination of two power laws with $P_{k}$ determining the rigidity where the transition occur, and the value of $c_{1}$ determines the slope of the power law below $P_{k}$. Its rigidity dependence and how it changes with time are shown in Figure \ref{fig5} as the upper set of graphs. 

We assume that $K_{\perp r}$ scales spatially similar to Equation (\ref{Eq16}) but with a different rigidity dependence at higher rigidity by simply replacing $c_{2 \parallel}$ with $c_{2 \perp r}$, which has a different value, to obtain $\acute{(K)_{\parallel}}$; we then assume
\begin{equation}
K_{\perp r} = 0.02 \acute{(K)_{\parallel}}
\label{Eq17}
\end{equation}
which is widely used and reasonable assumption, e.g. \citet{1999ApJ...520..204G}. 
On the other hand, $K_{\perp \theta}$, is more complicated, with consensus that $K_{\perp \theta}$ $>$ $K_{\perp r}$ away from the equatorial regions as discussed and motivated by \citet{2000JGR...10518295P} and \citet{2014SoPh..289..391P}. In this study, $K_{\perp \theta}$ is assumed to scale spatially similar to Equation (\ref {Eq16}) but again with a different rigidity dependence at higher rigidity obtained by replacing $c_{2 \parallel}$ with ${c_{2 \perp \theta}}$, which also has a different value than ${c_{2 \perp r}}$, to obtain $\grave{(K)_{\parallel}}$.
It is then assumed to be given by
\begin{equation}
K_{\perp \theta} = 0.02 f_{\perp \theta} \grave{(K)_{\parallel}} 
\label{Eq18}
\end{equation}
with
\begin{equation}
f_{\perp \theta} = A^{+} \mp A^{-} \tanh[8(\theta_{A} - 90^{\circ}) \pm \theta_{F}].
\label{Eq19}
\end{equation}
Here, A$^{\pm} = (d_{\perp \theta} \pm 1)/2 $, $\theta_{F}$ = 35$^{\circ}$, $\theta_{A}$ = $\theta$ for $\theta \leq 90^{\circ}$ but  $\theta_{A}$ = 180$^{\circ}$ - $\theta$ with $\theta \geq$  90$^{\circ}$ and $d_{\perp \theta}$ = 6.0 for protons and antiprotons.
This means that $K_{\perp \theta}$ has a different latitudinal dependence than the other diffusion coefficients which can be enhanced towards the heliospheric poles by a factor $d_{\perp \theta}$ with respect to the value of $K_{\perp r}$ in the equatorial region of the heliosphere. We keep $d_{\perp \theta}$ the same for the whole simulation period. The rigidity dependence and how it changes with time for these two MFPs are also shown in Figure \ref{fig5} as the middle group/band of curves. Note that in the equatorial plane $K_{\perp \theta}$ = $K_{\perp r}$ but not elsewhere. 
For a detailed theoretical discussion of these aspects of the diffusion tensor and theory, see the extensive reviews by \citet{2009ASSL..362.....S,2020SSRv..216...23S,2022SSRv..218...33E}. 
For a motivation and various applications of this particular modeling approach, see \citet{2000JGR...10518295P}, \citet{2003AnGeo..21.1359F}, \citet{2005AdSpR..35..554P,2012AdSpR..49.1660N}, \citet{2016AdSpR..58..453N}, to mention only a few. 

Focusing on the time variation (July 2006 - November 2019) for a specified rigidity (1.0 - 1.16 GV) of $\lambda_{\parallel}$, $\lambda_{\perp r}$ and $\lambda_{\perp \theta}$ as plotted in Figure \ref{fig4}(b), it is noticed that
they follow the exact same time variation, at this rigidity. However, the interest is on the trends over the solar cycle, and it is found that they have to increase gradually from 2006 - 2009 in order to reproduce the observations of this period. After December 2009, they keep reducing gradually as solar activity increases to reach the lowest values in February - March 2014, but in March - April 2015 a second short period of low values occur. This again is related to the rapid changes displayed for this particular period in Figure \ref{fig1}. 
From this time forward, they increase systematically towards the next solar minimum.
\subsection{Flux ratios over a solar cycle} \label{sec:ratio}
\begin{figure*}
\gridline{\fig{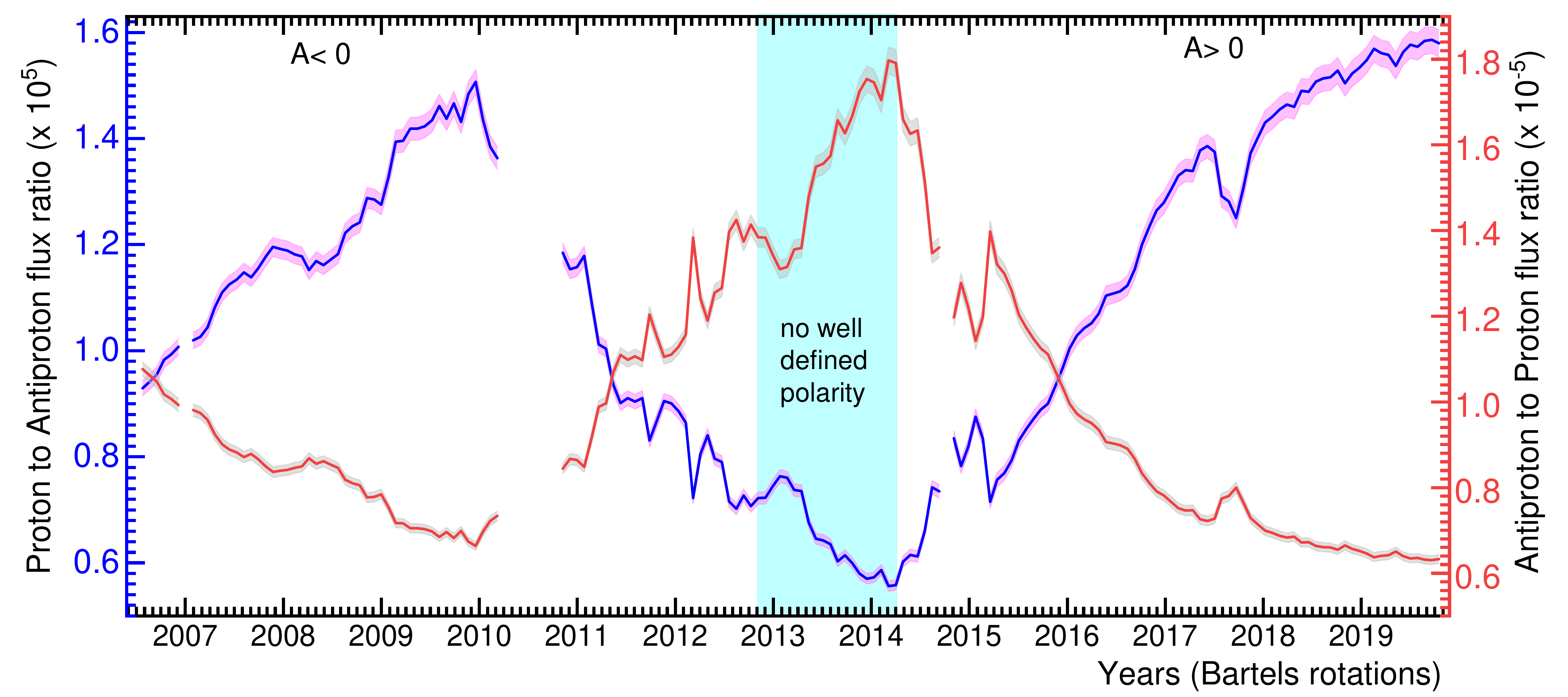}{0.95\textwidth}{}}
\caption{Variation of proton to antiproton flux ratio over time, from July 2006 - November 2019 (BRs 2361-2540) for 1.0 - 1.16 GV, shown in blue (left side). The magnitude is in the order of $10^{5}$. For convenience, the antiproton to proton flux ratio is shown in red (right side) with the magnitude in the order of $10^{-5}$. 
\label{fig6}}
\end{figure*}

The time variation of the antiproton to proton flux ratio at 1.0 - 1.16 GV for the simulation period is shown in Figure \ref{fig6}; it is found to be of the order of 10$^{-5}$. In the same panel, the proton to antiproton flux ratio is also shown. 
Noteworthy trends from July 2006 - December 2009 are that the latter ratio increased to a maximum value, after January 2010 it keeps increasing to reach a minimum value at the end of the polarity reversal; after the reversal, it increases rapidly within an year by $\sim$36\% with respect to February - March 2014. The changes at the end of 2014 and beginning of 2015 show a large fluctuation related to the changes in the flux. It then increases continuously, and more gradual, to reach again a high value when solar minimum sets in, with the exception of changing rapidly in the second half of 2017. These fluctuations correspond to what is mentioned above and related to what is shown in Figures \ref{fig1} and \ref{fig3}. 

\begin{figure*}[ht!]
\plotone{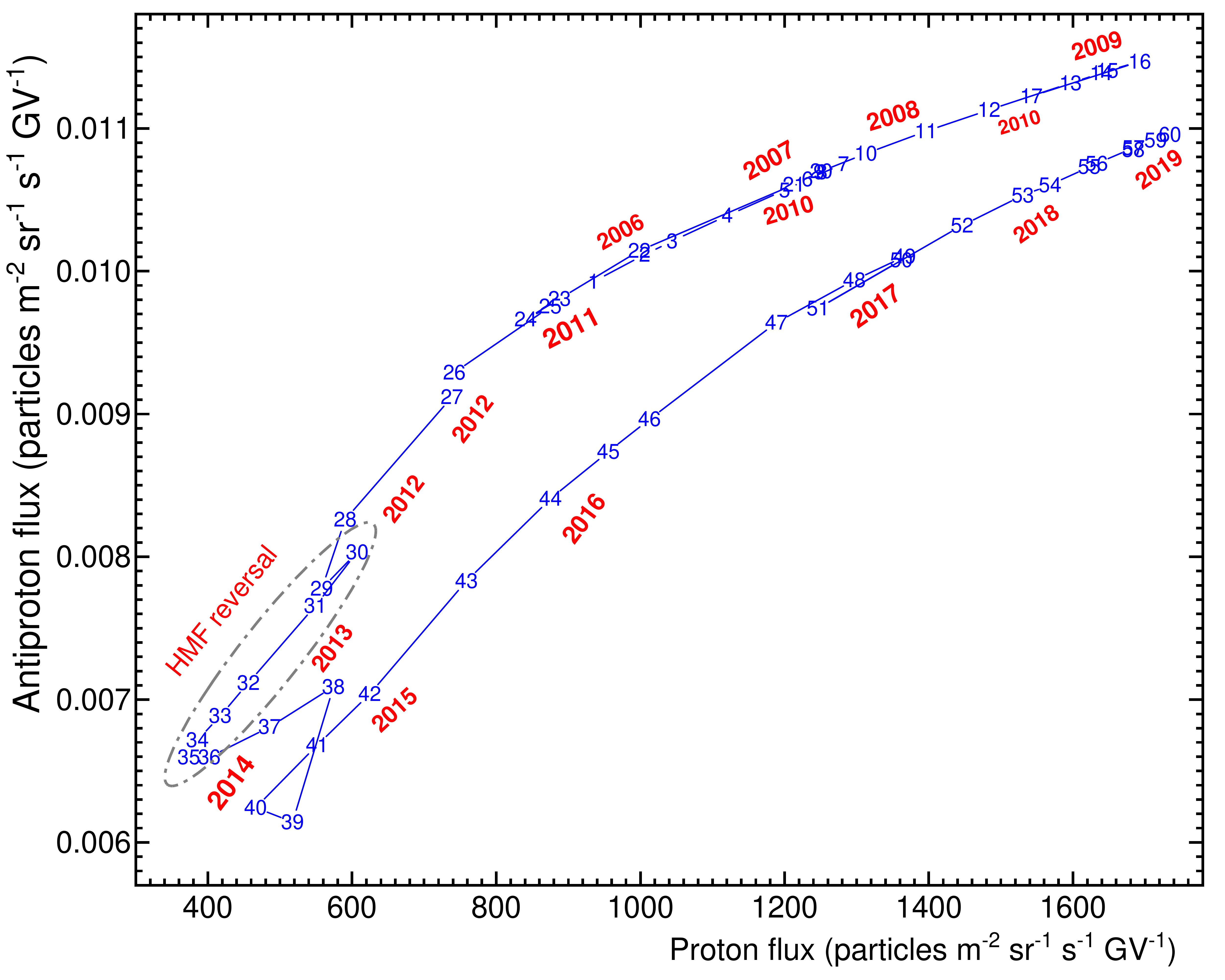}
\caption{
Hysteresis-like loop between antiproton and proton flux for 1.0-1.16 GV from July 2006 - November 2019 (BRs 2361-2540). Numbers (in blue) indicate 3-BR-averages, from "1" corresponding to BRs 2361-2363 and ends at "60" corresponding to BRs 2538-2540. Corresponding year numbers are marked in red. HMF reversal period is indicated with a looped grey dashed line.
\label{fig7}}
\end{figure*}

Apart from showing flux ratios, another interesting and insightful way to illustrate how differently the antiproton flux behaves with respect to the proton flux over the simulated period, is to produce a plot as shown in Figure \ref{fig7}. The obtained loop is popularly called a hysteresis behaviour. 
The numbers indicate 3-BR-averages, where "1" corresponds to July - September 2006, then e.g., "16" corresponds to November 2009 - January 2010, "29-35" to November 2012 - March 2014, which is the polarity reversal phase, and "60" corresponds to the September-November 2019, the end of the simulation period. Year numbers are also indicated to guide the eye, showing how the looped-curve increases from July 2006 to December 2009 with decreasing solar activity, decreases along the same slope up to the end of 2011 with increasing solar activity. Then in 2012 the shape (slope) begins to change, differently in the HMF reversal phase. Even after the reversal, the shape remains scattered for $\sim$1 year, then it begins to increase systematically up to the end of the simulation period but with the slope being somewhat different from 2017 onward, and almost the same at in the previous solar minimum.
\section{Discussions} \label{sec:discussions}
Concerning solar activity, the recent solar minimum between Solar Cycles 24 and 25 seems comparable with the previous solar minimum between Solar Cycles 23 and 24. The HCS tilt angle $\alpha$ is below 10$^{\circ}$ with less fluctuations since beginning of 2018, but the HMF magnitude at the Earth is higher than 4 nT and still having large fluctuations until the end of the simulation period. 
The lowest value for $\alpha$ recorded during the previous minimum was 4.5$^{\circ}$ in September-October 2009, with the field magnitude at 3.5 nT in March - April 2009. After this minimum both start to increase rapidly but with different rates, with $\alpha$ reaching the highest value 
in January - February, 2013. Afterwards it shows a very large fluctuation in 2014, and then progressively decreases with modest fluctuations. The HMF rate on the other hand flattens off during the polarity reversal period to reach its highest value of 7.9 nT afterwards, in November - December 2014. Then it decreases progressively with relative large fluctuations. According to \url{http://wso.stanford.edu/}, not shown in Figure \ref{fig1}, $\alpha$ reaches a minimum of 2.1$^{\circ}$ by March - April 2020 before starting to increase, with field magnitude reaching a minimum of 3.8 nT, also in March-April 2020. It is  noted that the Oulu Neutron Monitor recorded its peak count rate in mid-2020, and is observably lower than the counts recorded in the end-2009, see \url{https://cosmicrays.oulu.fi/}. This particular phenomenon where the flux of GCRs at Neutron Monitor energies (high rigidity) is lower in $A>0$ polarity cycles than in $A<0$ cycles, in contrast to what happens at low rigidity as discussed next, is explained in detail by \citet{2021AdSpR..68.2953K}; see also references therein.

Concerning the proton flux variations at 1.0 - 1.16 GV, it is noted that the flux gradually increases since July 2006 with the highest flux during 07 December 2009-03 January 2010 (BR 2407/CR2091). It then decreases as solar activity level increases and reaches the lowest flux level during 06 March - 02 April, 2014 (BR 2464), around the end of the polarity reversal phase. Afterwards, the flux starts to recovery and by mid-2019 it reaches almost the same level than in December 2009. Since protons drift inward mostly through the heliospheric polar region in an $A>0$ cycle, it is expected that during the recent minimum, somewhat higher fluxes than during the previous 2009 minimum should be observed at lower rigidity, given that all the heliospheric conditions are similar; see predictions by \citet{2014SoPh..289.3197S,2017A&A...601A..23P,2021AdSpR..68.2953K} and the recent report of proton spectra observed between 50-250 MeV in 2020 to be higher than in 2009 \citep{2023ApJL..00.0000M}. 

The best-known prediction of drift models with a changing HCS \citep{1981ApJ...243.1115J} is that the proton flux should exhibit a peaked intensity-time profile during $A<0$ polarity cycles, but flatter profiles during $A>0$ cycles, also evident from Figure \ref{fig3}; see also e.g., \cite{2003AnGeo..21.1359F,2021ApJ...908..167E}. It implies that the antiproton intensity-time profiles should be different than for protons, as is found and reported here but it appears not as spectacularly large as what the original drift or drift dominated modulation models predicted, e.g., by \citet{1979ApJ...234..384J,1985ApJ...294..425P,1995ApJ...442..847L,2004AdSpR..34..115F,2017ApJ...849L..32T} and references there-in.

Concerning flux ratios, it is further noted that the $\sim$ 60\% increase of the proton to antiproton ratio by 2009 indicates that even when long periods of low activity levels and quiet conditions occur, a significant rise in the antiproton flux does not consequently follow. The simulated antiproton to proton flux ratio in 2019 is lower than in 2009. The lowest proton to antiproton flux ratio is found in February-March 2014, although the lowest levels of the individual fluxes are almost a year apart. The trends in this flux ratio is qualitatively similar to the positron to electron ratio but with quantitative differences as simulated by 
\citet{2021ApJ...909..215A,2022arXiv221213397A}; 
see also \cite{2020JPhCS1690a2004M}.
Concerning the loop-behaviour between the proton and antiproton fluxes with time shown in Figure \ref{fig7}, it is emphasized that apart from the different VLIS's for these GCRs, the only difference between their modulation is their oppositely directed drift directions under similar modulation conditions. In this context, taken into consideration what is shown in Figure \ref{fig2} and together with the different drift patterns, the loop behaviour reflects exactly how differently these GCR particles is modulated. The changes in the slopes are significant from before to after the reversal period as the new drift directions settle in. After the reversal the loop is not repeating its 'path' as before but follows a new one to end up being lower in 2019 than in 2009 and with a somewhat different slope. This relates directly to the discussion above about the different behaviour during the two polarity cycles of oppositely charge particles.

An interesting trend in fluxes is evident during late-2017 when protons display a large and relatively long decrease, interrupting the recovery to solar minimum levels, and followed by a systematic recovery. Our simulated antiproton flux also shows this transient type behaviour but with quantitative differences which is considered as indicative of charge-sign dependence. This aspect, specifically what causes this type of transient, needs further investigation. It is probably an indication of the effect of a merged interaction region \citep{1998SSRv...83...33M,1998SSRv...83..309W} far beyond the Earth; see simulations by e.g., \cite{1999AdSpR..23..501L,2019ApJ...878....6L}.
\section{Summary and Conclusions} \label{sec:summary}

The main focus of this study is to improve our understanding of how antiprotons are modulated over a solar cycle. Comparing the simulated proton and antiproton spectra as well as the flux ratios give insight about charge-sign dependence over such a long period. Our simulations are validated by using observational proton spectra from PAMELA for July 2006 - May 2011 and from AMS for May 2011-November 2019, thus covering the solar minimum period before 2009, the period of maximum solar modulation including the time of the reversal of the HMF polarity, and then up to solar minimum conditions in 2019. 
It has turned out that the recent solar activity minimum around 2020 is comparable with the minimum between Solar Cycles 23 and 24 in the sense that the Sun was again rather quiet, also in terms of the tilt $\alpha$ of HCS, the magnitude of the HMF and the GCR spectra that are observed. See the recent reports of proton spectra observed at lower rigidity before and during 2020 by \cite{2020ApJ...901....8B,2023ApJL..00.0000M}.

The spectral shape and values of the VLIS's for various GCRs are an important aspect of modelling studies. In this context, the proton VLIS is well-established, perhaps as good as it can be done using models compared to observations beyond the HP and at the Earth at very high rigidity. The VLIS for antiprotons is not well-established, although quite reasonably determined by using GALPROP models and observations at high rigidity at the Earth. It is shown here that its rather peculiar VLIS plays an important role in the total amount of antiproton modulation between the HP and the Earth and in how its modulation develops over a solar cycle with respect to that for protons. However, comparing our modeling results with what was reported by \cite{1989ApJ...344..779W,1999PhRvL..83..674B}, it is evident that we have indeed make good progress, both in terms of what is assumed for the antiproton VLIS and what is found for the flux ratios over the reversal period of the HMF direction. Based on the newly determined VLIS for antiprotons, we found that the total modulation obtained with the model between the HP (122 au) and Earth (1 au) is vastly different than for protons. The peak intensity in the antiproton VLIS occurs at a much higher rigidity than for protons, subsequently also for the modulated spectra. The total modulation that antiprotons experienced at low rigidity is markedly less than for protons. 

A plot between antiproton and proton fluxes over the 2006 - 2019 period shows a clear hysteresis behaviour indicative of how differently anti-protons are modulated from protons over a solar cycle. Together with observations these computational results clarify to a large extent the phenomenon of charge-sign dependence in heliospheric modulation over a solar cycle. Our simulations show that even for the quietest heliospheric conditions not significantly more antiprotons, especially at lower rigidity, is reaching the Earth in sharp contrast to protons. 

Our results demonstrate how the underlying physics in the TPE varies over a solar cycle displayed by the computed time-variations of the main diffusion coefficients. In addition we found that the drift scale over time needs to be adjusted significantly with solar activity, with its lowest level during the phase of no well-defined HMF polarity. Furthermore, after the reversal period its recovery to large values associated with low solar activity is a slow process, it takes at least 2 years. This is consistent to what we reported when modeling electron and positron spectra over a similar period although the flux ratio of these particles is quantitatively different compared to protons and antiprotons.
         
\begin{acknowledgments}

The authors thank Mirko Boezio, Riccardo Munini and Nadir Marcelli for discussions of the PAMELA data and making the proton data set electronically available. The authors also wish to thank the GALPROP developers and their funding bodies for access to and use of the GALPROPWebRun service. We acknowledge the use of HCS tilt angles data from the Wilcox Solar Observatory 
(\url{http://wso.stanford.edu}), and HMF  data from NASA's OMNIWEB data interfaces (\url{http://omniweb.gsfc.nasa.gov}). 
This  work is supported by the Shandong Institute of Advanced Technology (SDIAT) and NSFC Project U2106201. MDN acknowledges the SA National Research Foundation (NRF) for partial financial support under the Joint Science and Technology Research Collaboration between SA and Russia (Grant no:118915).

\end{acknowledgments}



\end{CJK*}
\end{document}